# The monochromator beamline at FLASH: performance, capabilities and upgrade plans

Natalia Gerasimova[*], Siarhei Dziarzhytski and Josef Feldhaus

DESY, Hamburg, Germany

The monochromator beamline at the FLASH facility at DESY is the worldwide first XUV monochromator beamline operational on a free electron laser (FEL) source. Being a single-user machine, FLASH demands a high flexibility of the instrumentation to fulfil the needs of diverse experiments performed by a multidisciplinary user community. Thus, the beamline has not only been used for high-resolution spectroscopy that it was originally designed for, but also for pump-probe experiments controlling the temporal-spectral properties at moderate resolution, and as a filter for high harmonics of the FEL at very low resolution. The present performance and capabilities of the beamline are discussed with emphasis on particularities arising from the nature of the FEL source, and current developments are presented aiming to enhance its capabilities for accommodating a wide variety of experiments.

Keywords: FEL, XUV, monochromator, diffraction gratings, femtosecond pulses

## 1. Introduction

FLASH (Free electron LASer in Hamburg) is a single-pass free electron laser (FEL) based on self-amplified spontaneous emission (SASE) (1, 2). Since 2005 FLASH, the worldwide first FEL working in the extended ultraviolet (XUV) range, has been operated as a user facility. FLASH provides pulses of coherent radiation with pulse durations down to femtoseconds and pulse energies of tens to hundreds of microjoules. The wavelength of the fundamental falls in the range 4.2 – 47 nm, the typical bandwidth is 1%. Higher frequency harmonics extend the wavelength range to below 1 nm. Up to now FLASH has been a single-user facility allowing only a limited number of experiments. At the same time, the diversity of the pioneering experiments that were chosen to be performed is very large, ranging from atomic physics through materials science to biological applications, and the experiments diverge not only with

---

[*] Corresponding author. Email: natalia.gerasimova@desy.de

respect to the materials investigated but also with regard to the applied techniques. Such versatility demands highly flexible instrumentation and beamlines. The diversity of user needs is much wider than usual for a synchrotron beamline. In addition to standard parameters like resolving power and throughput, the timing structure became of importance for certain applications. Thus, the monochromator beamline at FLASH exploits the full range of its capabilities from an ultimately high resolution to a maximum throughput mode of operation. From the point of view of the instrument one can sort the experiments performed at this beamline in the following way: (i) High resolution applications. These experiments profit from the ultimately high brilliance of FLASH which enables for the first time "photon hungry" experiments demanding a high resolution in the XUV range. Examples are low-cross section experiments like inelastic scattering or resonant fluorescence spectroscopy of highly charged ions (3). The average brilliance of the source is of main importance for most of such applications, therefore high repetition rate FELs such as FLASH which is based on superconducting accelerator technology, are the proper choice. At high resolution, however, the FEL pulses become elongated to the picosecond range. (ii) Experiments demanding a moderate resolution, but better than the inherent FEL bandwidth, mainly for time-resolved studies using pump-probe techniques. Time- and angle-resolved photoemission spectroscopy is one of the applications aiming to control the temporal-spectral resolution of the monochromator. (iii) Experiments demanding extremely high throughput and minimal temporal elongation of pulses (extremely low resolution) in order to provide filtering of higher frequency harmonics of the FEL (see for example the coherent magnetic scattering experiment using the $5^{th}$ harmonic of the FLASH radiation (4))

The present paper discusses the actual status of the monochromator beamline at FLASH with emphasis on particularities arising from the properties of the SASE FEL source, e.g. temporal and statistical properties, and presents current developments aiming to further enhance its capabilities for accommodating a wide variety of experiments.

**2. Beamline overview**

The monochromator is a plane grating monochromator of SX-700 type proposed by Petersen (5). The SX-700 mount is characterized by a fixed focus constant,

$c_{ff}=\cos\beta/\cos\alpha$, where $\beta$ is the angle of diffraction and $\alpha$ is the angle of incidence on the grating. Both the source point and the exit slit position are fixed in space for all photon energies when the cff-value is fixed. The introduction of a plane mirror positioned before the grating allows to vary the included angle on the grating and thus to keep the cff value fixed when the wavelength is changed, so that the geometry of the beamline before and after the mirror - grating pair stays unchanged. In addition, the concept of operating in collimated light allows keeping the image point fixed for different cff values (6). This design has become very popular in the soft X-ray regime since it provides high flexibility and allows a free choice of compromise between flux and resolution while working with the same grating.

Details of the design of the monochromator beamline at FLASH can be found in (7). The beamline comprises two branches: PG1 and PG2. The PG1 branch is permanently equipped with a double Raman spectrometer and the PG2 branch provides beam for user experiments of different types. The Raman spectrometer has been designed for high-resolution resonant inelastic scattering experiments and is currently in the commissioning phase. The PG2 branch has been operational for users since 2005.

We will now present the different modes of monochromator operation and their key parameters.

### 2.1. High resolution mode

The monochromator has been designed to be capable of providing a resolving power of 10.000 – 70.000 depending on photon energy range and grating in use (7).

The standard method for characterizing the resolution of synchrotron XUV beamlines is based on photoionization or absorption measurements behind the monochromator exit slit. In the case of a SASE FEL light amplification starts from shot noise in the electron beam, thus shot-to-shot fluctuations in the spectral distribution are inherent to FLASH radiation (8). Consequently, scanning of the monochromator over a resonance appears to be more complicated for a SASE FEL source than for a synchrotron beamline, and demands stable SASE operation during the scanning time (i.e. long-term stability) and / or reference intensity measurements behind the exit slit. While the applicability of photoionization measurements behind the monochromotor exit slit at FLASH is currently under investigation and optimization, another method for resolution measurements has been used extensively. The method is based on absorption measurements with the monochromator operated in the spectrometer mode. In this way scanning is avoided, and the information on the whole spectral region of the resonance can be acquired on a single shot basis. This type of measurements does not take into account the influence of an exit slit. The spectrometer mode has the advantage that data acquisition is less time-consuming than in the monochromator mode; in our case less than 5 minutes acquisition time were needed to measure the average spectral distribution.

For the absorption measurements of the PG2 beamline in spectrometer mode (9) the gas filled attenuator (a 15m long windowless volume located between the undulator and the first optical element (10)) was filled with Ne (typical pressure $1.5*10^{-2}$ mbar). A fluorescent crystal (yttrium aluminum garnet (YAG) doped with 0.2% Ce, 0.2 mm thick, produced by Crytur Ltd., Czech Republic) was placed in the exit slit position. The spectral distribution was registered by an intensified CCD (Andor, model iStar DH740). A typical spectrum resulting from the radiation passing through Ne gas around 48eV photon energy is presented in Fig.1a. The spectrum in this range is characterized by the singly excited Rydberg series $2s2p^6 np$ overlapping with weaker doubly excited Rydberg series $2s^2 2p^4 3snp$ (11). To estimate the resolution we analyse the measured width of the lines of the $2s2p^6 np$ series for n≥6. Since resonances for n≥6 are characterized by widths below 1.3 meV we attribute the measured width of several meV to the resolution of the monochromator (without applying a deconvolution procedure). The resolution of the 200 l/mm grating operating in first diffraction order with cff=3 was estimated in this way; the result is presented in Fig.1b. The resulting resolution of about 4.2 meV corresponds to a resolving power of better than 11.000, while ~20.000 is expected from the design.

Investigating the dependence of the resolution properties on the distance of the YAG crystal from the grating allows us to suggest that the deterioration of the resolution is connected to the longitudinal properties of the source. First, the waist of the longitudinal resolution dependence corresponding to the focal plane differs for different lasing conditions, and for the case presented in Fig. 1b it is located about 30 mm upstream the nominal position. According to ray-tracing simulations using the Shadow code (12), this situation corresponds to an effective source located 8 m upstream the undulator exit. Indeed, the saturation length of the SASE process is expected to become smaller for longer wavelengths (for the presented case no care was taken to move the saturation point of the FEL to the undulator exit). Second, the measured longitudinal dependence is more shallow than predicted by ray-tracing for a source of no depth. Such behaviour can be explained by a linear combination of two effective sources located at the end and deep in the undulator, as expected since for a SASE FEL the process of field growth does not stop at the saturation point due to the spectral broadening (8). In other words, the over-saturation is not an optimal mode of SASE FEL operation for experiments demanding high resolution since the source properties can degrade the resolving power of the monochromator.

The measured resolving power better then 11.000 for the 200 l/mm grating confirms the high performance of the beamline in terms of resolution. Nevertheless the 200 l/mm grating is not optimized in terms of efficiency for high-resolution operation; it is rather a "moderate resolution" grating. The calculated efficiency curves of the presently installed 200 l/mm and 1200 l/mm gratings are shown in Fig.2a and Fig.2b, respectively; restrictions on the photon energy range for particular grating / cff values come from mechanical constraints and the limitation of the glancing angle of incidence on optical elements to be below 6° to avoid FEL induced damage of optical surfaces. The calculation has been done using the "Reflec" code (13) included in the "Ray" package (14) developed at BESSY; the code uses the differential formalism developed by Neviere et al. (15). The 200 l/mm grating operating at low cff values close to 1.5 allows moderate resolution mode in the whole range of the FLASH fundamental with high efficiency. At the same time the high resolution mode (the 200 l/mm grating set to high cff for the 30 – 80 eV photon energy range and the 1200 l/mm grating for the 70 – 250 eV range) is available only with considerably lower flux. The high resolution mode is presently used for diagnostic of the SASE FEL radiation properties when the beamline operates in the spectrometer mode. On the other hand, experiments demanding high resolution at FLASH are mainly "photon hungry" experiments, thus flux is an essential parameter. In order to provide high resolution capabilities along with high efficiency over the full range of the FLASH fundamental, a new dedicated grating has been designed. A groove density of 600 l/mm is optimal to fit the 30 – 250 eV working range with cff set to 3; the choice of these parameters allows a resolving power above 20.000 over the whole photon energy range of the FLASH fundamental. The optimization of the laminar groove profiles (optimum depth of grooves: 30 nm, valley to spacing ratio:

0.65) for a C-coated 600 l/mm grating results in efficiencies presented in Fig.2c, so that a substantial increase of efficiency for the high resolution mode is expected.

*2.2. Controlling temporal-spectral properties*

Gratings are known to stretch ultrafast pulses. An estimation of pulse stretching induced by the gratings has been done based on geometrical optics considerations. The path length difference in the dispersion direction causes the pulse stretching: $T=(N*m*\lambda)/c$, where $N$ is the number of illuminated grooves, $m$ the diffraction order, $\lambda$ the wavelength, $c$ the velocity of light. The number of illuminated grooves depends on the groove density $k$ and the beam cross-section in the dispersion direction $a$: $N=(a*k)/\cos\alpha$. For an estimate of the beam cross-section $a \approx \sigma*d$, the distance $d = 49$m between the undulator exit and the collimating toroidal mirror prior to the monochromator and the $\sim\sqrt{1/\hbar\omega}$ dependence for the divergence $\sigma$ of the FEL beam in the far zone have been used. Both the longitudinal source position and the beam divergence depend on the lasing conditions; such variations have not been taking into account. For the particular lasing conditions the beam cross section $a$ can be measured using imaging screen located 54 m downstream the undulator exit, behind the collimating mirror.

The calculated stretching of FLASH pulses for the 200 l/mm and the 1200 l/mm gratings, as well as for the planned 600 l/mm grating is shown in Fig. 3a. Typically the high resolution mode (i.e. high cff, high groove density) produces stretching to picoseconds, and stretching is more pronounced at low photon energies. On the other hand, the moderate resolution mode (i.e. low cff value, low groove density) allows to keep the pulse duration within hundreds of femtoseconds and below. Thus, the moderate resolution mode is of interest for techniques like trARPES (time-resolved angle-resolved photoemission spectroscopy) demanding a compromise between temporal and spectral resolution. In addition to changing monochromator settings one can confine the number of illuminated grooves by closing an aperture located in front of the monochromator. Diminishing the pulse elongation costs flux in this case; besides, this possibility is limited by diffraction effects and is currently under investigation.

*2.3. Monochromator as a filter for high FEL harmonics*

Radiation of a SASE FEL in the saturation regime contains higher frequency harmonics. A contribution of third and fifth harmonics to the total output energy of 0.6% and 0.03%, respectively, has been reported for the FLASH high performance operation at 13.7 nm wavelength (1). The high harmonics radiation can be harmful for certain experiments or may be used for other experiments. The high harmonics including $3^{rd}$ and $5^{th}$, expand the FLASH working range to 1.4 keV and comprise: (i) the so-called "water window", namely the range between the C K-edge and O K-edge relevant for a number of biological applications; (ii) the oxygen K-edge and (iii) the L-edges of the 3d elements such as Fe, Co, Ni, etc. As this photon energy range is highly interesting and demanded for experiments, the use of the monochromator beamline as a filter for high FEL harmonics has been considered. Filtering can be done by other methods as well, e.g. foils are known to transmit photons within a certain range of interest and attenuate within a harmful region. However, the attenuation ratio for the energy spacing between $1^{st}$ and $3^{rd}$ harmonics and even $1^{st}$ and $5^{th}$ harmonics is moderate for the foils and achieving a higher attenuation ratio is accompanied by substantial reduction of flux. Care must be taken to avoid damage of the foils and coherent scattering from the foils should be taken into account. The advantage of a monochromator to be used as a filter is the clean spectrum it provides.

Applications using high harmonics demand mainly high flux and preservation of the ultra-fast pulse structure. For this case a monochromator with the smallest cff values along with minimal groove density ensures minimal temporal pulse stretching. The same choice of parameters results in smallest dispersion as well and, taking into account the large FEL bandwidth and the 1:1 imaging of the exit slit onto the sample in dispersion direction, in high photon density on the sample. Presently the 200 l/mm grating with cff set to 1.1 – 1.3 is used for high harmonics filtering. These settings result in a grating efficiency of 15% deep in the water window, but the corresponding total PG2 beamline transmission is only 2% (Fig.4a). The limitations are coming from the fact that all optical elements in the beamline are coated with diamond-like amorphous carbon (16, 17). This coating ensures high damage resistance and excellent beamline transmission below 250 eV (the design was made for the 20 – 200 eV working range expected for the FLASH fundamental harmonic) but restricts the higher photon energy range to below 900 eV and allows only moderate transmission above the C K-edge. In addition the 200 l/mm grating has not been optimized for the high energy range and its efficiency above 550 eV is below 2%.

In order to increase the capabilities of the beamline, a dedicated grating to be used as a filter for high harmonics has been designed. The minimal grove density available on the market, 80 l/mm, has been chosen. Such a groove density allows low cff values of 1.1 – 1.2 to be used in the range of 300 – 900 eV preserving the pulse length within tens of femtoseconds (Fig. 3b). For the beamline geometry optimized for high resolution mode, even lowest cff values close to 1 result in glancing angles on the grating above 3°. For such glancing angles high reflectivity in the range 300 – 900 eV can be achieved by using high Z element coatings, therefore platinum has been chosen for this grating. On the other hand, high Z element coatings are expected to be less damage resistant than coatings of low Z elements. To avoid damage, the grating is to be used in combination with gas or foils attenuating the high power of the fundamental. To cover the whole range of 300-850 eV with high efficiency, a solution proposed by Jobin-Yvon, namely variable groove depth gratings, was chosen. The gradual change of depth increases the range of high efficiency if the grating is sufficiently wider than the beam footprint. Results of the laminar groove profile optimization are presented in Fig. 4b; the depth of grooves changes from $h_{min}$ = 7 nm to $h_{max}$ = 15 nm, the valley to spacing ratio is 0.55. Using this proposed grating, the transmission of the beamline in the range of 600 – 900 eV would increase by more than an order of magnitude compared to the present status.

## 3. Conclusions

The monochromator beamline at FLASH allows high flexibility in operational modes from ultimately high resolution to ultimately high throughput depending on applications. The high level performance is confirmed, a resolving power better than 10.000 has been measured. On the other hand, the SASE FEL source properties can affect the beamline performance. Another particularity of the FEL compared to a synchrotron source is the femtosecond time structure; the grating effect on the pulse duration has to be taken into account. To increase the capabilities of the beamline, two new gratings, a high resolution grating for the FLASH fundamental range and a grating-filter for high harmonics, have been designed and are planned to be installed.

**Acknowledgments**

The authors thank H. Weigelt for technical support, T. Nunez and T. Kracht for assistance with the experimental control, and V. Rybnikov for the implementation of the data acquisition.

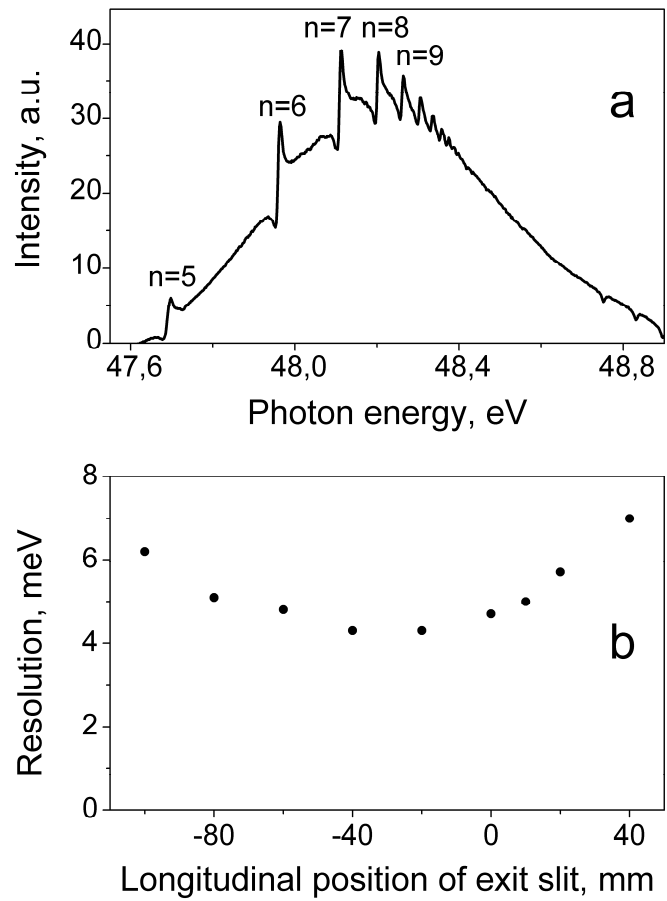

Figure 1. (a) Spectrum averaged over 1200 shots of FEL radiation after passing the Ne-filled gas attenuator. The spectrum was measured at the PG2 beamline operating in spectrometer mode using the 200 l/mm grating in first diffraction order with cff=3; (b) Resolution of the PG2 beamline as a function of the longitudinal exit slit position estimated from the measured widths of absorption lines (see text).

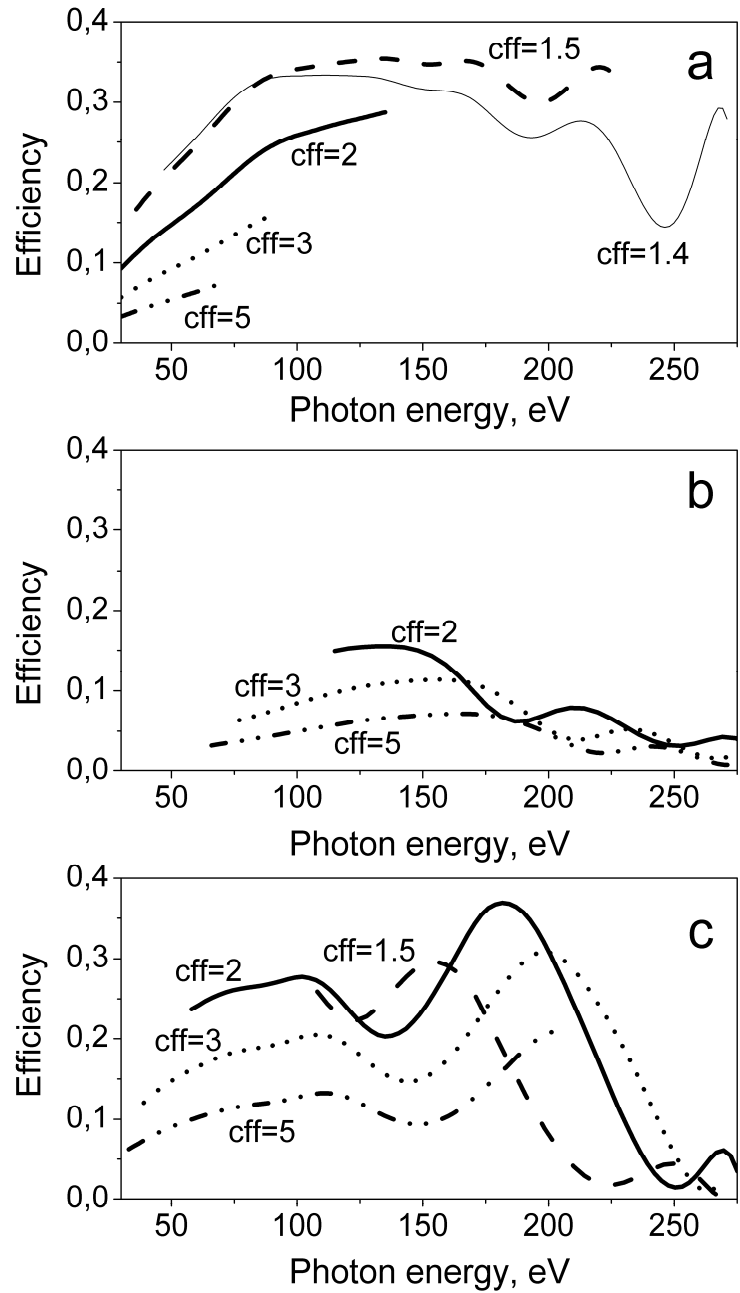

Figure 2. Calculated efficiency of the 1st diffraction order of (a) the present 200 l/mm grating, (b) the present 1200 l/mm grating, and (c) the newly designed 600 l/mm grating, for different cff values: cff=1.4 (thin solid line), cff=1.5 (dashed line), cff= 2 (thick solid line), cff=3 (dotted line), and cff=5 (dash-dot-dot line).

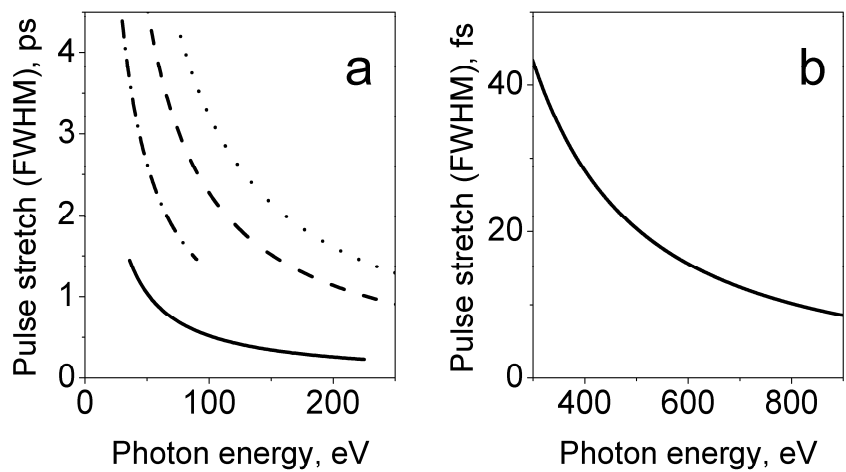

Figure 3. Calculated pulse stretching (FWHM) (a) in the range of the FLASH fundamental induced by the 200 l/mm grating (cff=1.5, solid line; cff=3, dash-dot-dot line), the new 600 l/mm grating (cff=3, dashed line) and the 1200 l/mm grating (cff=3, dotted line); and (b) pulse stretching induced by a 80 l/mm grating working as filter for high harmonics.

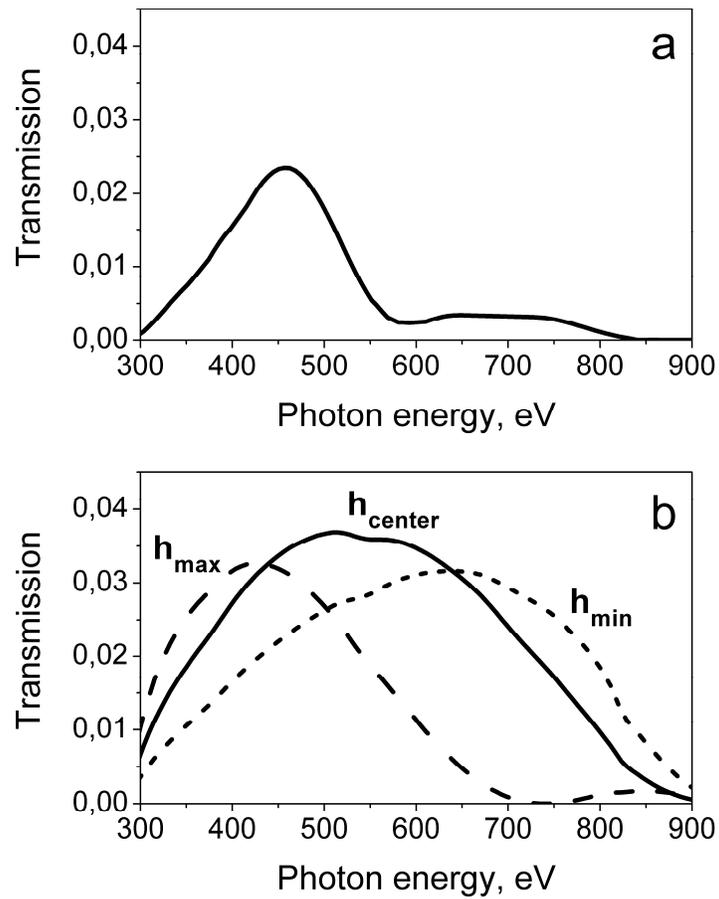

Figure 4. Calculated transmission of the PG2 beamline operating with (a) the 200 l/mm grating and (b) the newly designed 80 l/mm grating in the range of high harmonics. Cff values range from 1.1 to 1.3. The groove depths $h_{max}$, $h_{center}$ and $h_{min}$ correspond to different lateral positions on the grating (see text).